# Electron tomography for functional nanomaterials

Robert Hovden and David Muller

Robert Hovden, Department of Materials Science and Engineering, University of Michigan, USA; hovden@umich.edu

David Muller, School of Applied and Engineering Physics, and Kavli Institute for Nanoscale Science, Cornell University, USA; dm24@cornell.edu

Modern nanomaterials contain complexity that spans all three dimensions—from multi-gate semiconductors to clean energy nanocatalysts to complex block copolymers. For nanoscale characterization, it has been a longstanding goal to observe and quantify the three-dimensional (3D) structure—not just surfaces, but the entire internal volume and the chemical arrangement. Electron tomography estimates the complete 3D structure of nanomaterials from a series of two-dimensional projections taken across many viewing angles. Since its first introduction in 1968, electron tomography has progressed substantially in resolution, dose and chemical sensitivity. In particular, scanning transmission electron microscope tomography has greatly enhanced the study of 3D nanomaterials by providing quantifiable internal morphology and spectroscopic detection of elements. Combined with recent innovations in computational reconstruction algorithms and 3D visualization tools, scientists can interactively dissect volumetric representations and extract meaningful statistics of specimens. This article highlights the maturing field of electron tomography and the widening scientific applications that utilize 3D structural, chemical, and functional imaging at the nanometer and subnanometer length scales.

**Keywords**: electron tomography, scanning transmission electron microscopy, transmission electron microscopy, electron energy loss spectroscopy

**Electron tomography born**

Tomography, from the Greek words tomos (to slice) and graphe (to draw), describes a technique where materials are sectioned to reveal hidden internal structure. However, electron tomography does not measure specimen slices directly, but instead reconstructs the volumetric structure of nanomaterials from a set of high-resolution projection images across many viewing angles. Electron tomography is advantageous for measuring volumes in the range of $(1000 \text{ nm})^3$ to $(10 \text{ nm})^3$ at resolutions around 30 to 3 Å.[1] Precise and accurate three-dimensional (3D) reconstruction of nanoscale materials has had decades of advancements in electron microscopy, and data processing and visualization.

Understanding material structure in three dimensions is significant for both biological and inorganic systems—natural or synthetic. Despite the recent surge of interest, electron tomography remarkably predates the personal computer. In 1968, DeRosier and Klug famously reconstructed the helical structure of the T4 bacteriophage tail from a single transmission electron microscope (TEM) projection using prior knowledge about its helical symmetry.[2] In 1970, spherical symmetry was exploited to reconstruct a negatively stained human wart virus and tomato bushy virus from multiple projections.[3] Electron tomography, as we know it, was achieved with 3D reconstruction of the low-symmetry fatty acid synthetase molecule in 1974.[4,5] In the last two years (2018–2019), more biological structures have been reported by electron tomography than all of the previous years combined in the European Protein Data Bank.[6]

The demand for electron tomography of nanomaterials has now blossomed across many fields—such as semiconductors and clean energy nanocatalysts. Initially, biological TEM techniques were applied to reconstruct polymer morphologies where similar contrast mechanisms could be employed.[7] In 2000, TEM tomography of silver particles on zeolite opened up 3D characterization of inorganic matter.[8] While biological specimens are radiation sensitive ($<\sim 1$ e$^-$/Å$^2$),[9] many inorganics can withstand orders of magnitude higher beam dose (as high as $10^6$ e$^-$/Å$^2$),[10] allowing a wider range of characterization methods. Scanning transmission electron microscope (STEM) tomography of Pd–Ru bimetallic nanocatalysts on mesoporous silica made nanoscale tomography quantifiable for the first time.[11] Electron tomography of functional nanomaterials is advancing at a rapid pace, first with data digitization, but more recently, computational hardware/software capable of efficient 3D reconstruction and visualization, new reconstruction algorithms inspired by compressed sensing mathematics,[12] high-efficiency spectroscopy, and the maturation of STEM.[13] We are now pushing the limits of distinguishing chemistry, spatial resolution, and object sizes that are bounded by fundamental sampling and dose requirements. This article on STEM tomography highlights the science and technological advances occurring for 3D structural, chemical, and functional imaging at nanometer length scales.

## STEM tomography of nanomaterials

STEM tomography was a marked innovation for characterizing functional nanomaterials in three dimensions. Midgley et al. utilized annular dark-field (ADF) STEM to produce 3D reconstructions where the value of each voxel was proportional to the proton density that comprised the material.[14] A key feature of ADF-STEM is that it provides a simple projection to allow for quantifiable tomography. This so-called "Z-contrast" image is formed when elastically scattered electrons are collected in a high-angle annular dark-field detector (HAADF)—a Rutherford scattering process. ADF-STEM overcomes the limitations of TEM for nanomaterials where phase contrast reversals degrade the projection requirement of tomography. STEM tomography is also better suited for thicker or crystalline materials where projections are less influenced by phase contrast. In STEM, the lateral dimension of the converged electron beam sets the two-dimensional (2D) diffraction limited resolution around 1–3Å (two times better than TEM's Scherzer resolution for the same aperture cutoff) which can easily resolve nano- to atomic-scale features in each projection.

**Figure 1** shows the 3D structure of a $Co_2P$ hyperbranched particle[16] reconstructed from a sufficient number of STEM projections taken across the widest angular range possible. Using the simplest reconstruction method, each projection image is aligned and re-projected to create a 3D representation of the object.[15]

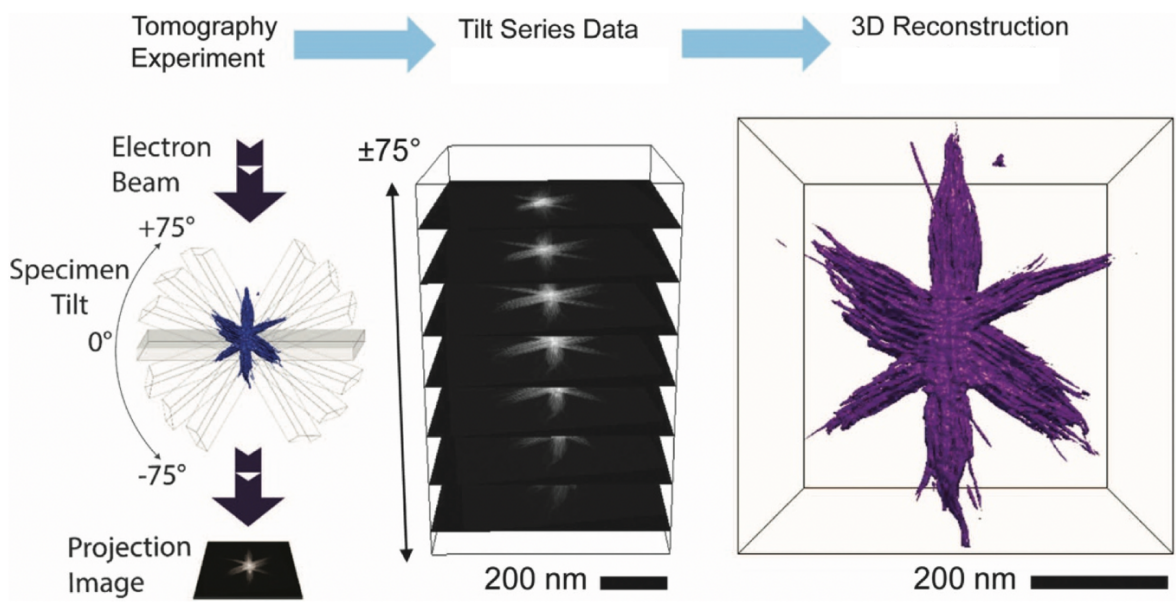

**Figure 1.** Illustration of the electron tomography data acquisition and reconstruction process. Two-dimensional projection images acquired of a three-dimensional (3D) object at different viewing angles comprise the raw experimental image stack (i.e., a tilt series). The tilt series is aligned, and a reconstruction algorithm is applied to produce a 3D reconstruction of the object. Here, a 3D isosurface visualization of a CoP$_2$ nanoparticle[1] is rendered using the tomography platform, tomviz.

*Materials application: Semiconductors*

The semiconductor industry's need for and spending relating to nanoscale metrology has continued to grow over the last 10 years. The International Technology Roadmap for Semiconductors (ITRS) specifically emphasizes a need for 3D characterization of transistors, memory, and interconnects.[17] This is motivated by the technological shift to stacked 3D device architectures that contain feature sizes at or below 1 nm. As a strategy to beat Moore's Law, the industry is leaving the confines of a 2D plane to pack more devices into a 3D volume. For computing, the FinFET, MuGFET, and gate-all-around transistor (GAA) are 3D electronic devices that demand electron tomography for characterization. Tomography of GAA wires showed noticeable interface roughness in the gate oxide layer (~3 nm HfO$_2$).[18] Three-dimensional memory architectures, such as Intel's and Micron's 3D-XPoint phase-change memory, are already commercially available with high-density, low-latency bit storage. In current phase-change memory, pinning due to device inhomogeneity causes partial or stuck phase switching that limits cycle life.[19] To diagnose device failure or degradation and improve the reliability of semiconductor devices, the entire 3D structure needs to be characterized with nanometer or even subnanometer resolution.

STEM tomography has been used to identify local defects in the vias of DRAM[20] and stress void formation in Ta-lined copper interconnects.[21] Many defects are localized in 3D space and are not quantifiable, or overlooked entirely, from a single projection image. Xin et al. have shown that low-*k* dielectric materials made from nanoscale pores with a subnanometer size distribution are below the detectible limit of ellipsometric porosimetry, necessitating high-resolution tomography to understand dielectric metamaterials.[22] Distinguishing chemistry within a material with ADF-STEM tomography is possible when constituents in a specimen have dissimilar elastic cross sections (normalized by their density). Markedly distinct species, such as the Ta ($Z = 73$) lining on a copper ($Z = 29$) interconnect, can be chemically characterized.[20,21] However, for more complex chemistries—such as thin $SiO_2$ oxide layers at a Si interface—knowing the exact chemistry in three dimensions is challenging. Current tomographic research is extending beyond HAADF to include spectroscopic techniques that provide chemical information.

**Seeing chemistry in 3D**

STEM allows chemically sensitive spectroscopic techniques, such as energy dispersive x-ray (EDX) and electron energy-loss spectroscopy (EELS), to be simultaneously acquired as the beam is scanned across a specimen. When combined with specimen tilt, these spectroscopic images can, in principle, allow for 3D tomographic spectroscopy.[23] However, only with recent advances in the useable beam current and detector sensitivity has spectroscopic tomography become viable.[24,25] In particular, energy-dispersive x-ray detectors (a silicon drift device) are now substantially larger (~170 $mm^2$) and closer to the specimen to achieve collection angles of ~1 steradian (~8% fractional area) corresponding to roughly a fourfold increase in efficiency.[24] Furthermore, multiple energy-dispersive x-ray spectroscopy (EDX) detectors (as many as four)[25] integrated around the specimen allow a high stage tilt range while preserving collection efficiency and ensuring predictable symmetry with tilt.[10] Lepinay et al. showed 28-nm transistor chemistry in three dimensions in a PMOS (**Figure 2**b) by using new EDX detector geometries and higher beam currents.[26] The work discerned seven elements within a complex metal gate stack containing just a few nanometers of TiN and $HfO_2$ oxide layers.

Chemical tomography has also been performed using EELS. An EELS spectrum measures all inelastic scattering processing, including excitations of core energy and valence electrons in a nanomaterial. Early demonstrations of chemical tomography used STEM-EELS to identify plasmon modes associated changes in the valence electron density.[27] An energy selecting filter was placed at the respective plasmon peak energies and the instrument was operated in TEM mode (i.e., energy-filtered TEM) as the specimen was tilted.[28] With this approach, materials indistinguishable by traditional electron tomography, such as $SiO_2$ in Si (Figure 2a)[27] and carbon nanotubes in nylon,[29] could be chemically mapped in three dimensions.

Plasmons offer large inelastic scattering cross sections to provide higher signal-to-noise (SNR) ratios in 3D reconstructions.[30] However, chemistry is only determined when the plasmon resonance energies are sufficiently distinct. More often, chemistry is extracted from the core electron excitations occurring at higher energies with smaller inelastic cross sections that sit upon a large background. Although EELS allows substantially higher collection efficiency (~90%) when compared to EDX (~8%), the SNR of core-loss excitations degrades rapidly from multiple inelastic scattering events in thicker specimens—especially problematic for lamellar samples at high tilt. In most cases, the SNR limits the quality and resolution of chemical tomography. EELS tomography of core-loss chemistry has been demonstrated in nanoparticles[31] and semiconductors.[32] Notably, in 2009, Si bonding of Si(Ti,N), Si, and Si(O) was tomographically imaged in three dimensions using EELS fine-structure that reflected differences in the local density of states.[32]

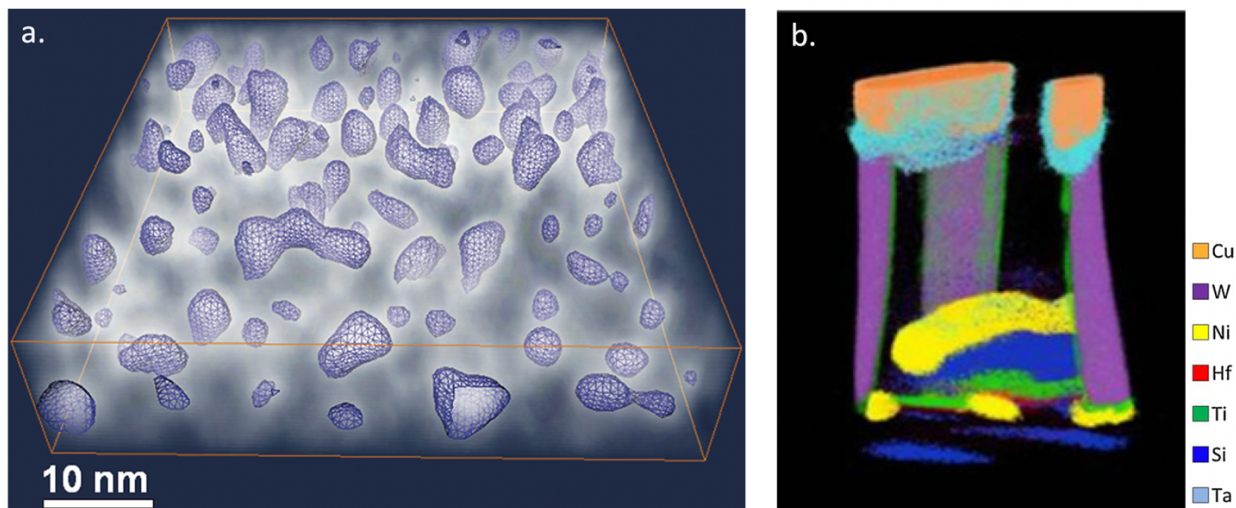

**Figure 2.** Chemical tomography of semiconductor materials. (a) Irregularly shaped silicon nanoparticles in a silica matrix reconstructed using plasmon resonance energies.[27] (b) 28-nm *p*-type channel metal oxide semiconductor transistor reconstructed using energy dispersive x-rays. Reprinted with permission from Reference 26. © 2013 Elsevier.

*Materials application: Electrochemistry and nanocatalysis*

Clean energy nanomaterials have performance intimately connected to chemical 3D structure—be it nanocatalysts for water splitting, hydrogen fuel cells, or electrodes on a battery. Surfaces where reactions occur often maximize the area through tortuous and porous morphology that is obfuscated in a 2D image. ADF-STEM tomography has revealed internal pore structure of a spongy $Cu_3Pt$ catalyst with enhanced activity in the oxygen reduction reaction of a fuel cell,[33] quantified the loading of Pt nanoparticles and their position on or within carbon supports in hydrogen fuel-cell vehicles,[34] and revealed their coarsening and lifetime degradation.[35] Quantitative analysis of 3D tomography can be used to extract the surface area of nanoparticles inside or outside support materials and correlate it to bulk electrochemical measures of performance (**Figure 3**). Spectroscopic tomography has begun investigating materials where nonhomogeneous chemical structure is optimized for catalysis, energy storage, or cost reduction. Genc et al. revealed the inhomogeneity of Ni in the popular lithium-ion cathode material, LiNiMnO.[36] Xia et al. correlated 3D elemental segregation in Ni-Fe nanoparticles and the formation of a hollow cavity structure to oxidation processes (**Figure 4**).[37]

Quantifying chemistry with STEM EDX and EELS is becoming possible across a wider range of functional nanomaterials. However, the high doses required by chemical tomography are often unachievable and limit the SNR. Worse, the limited dose also restricts the number of projections that can be acquired and this reduces the resolution and object size of the final 3D reconstruction. Thus, chemical STEM tomography of nanomaterials remains an active area for progress, as detector sensitivity and readout rates continue to improve.

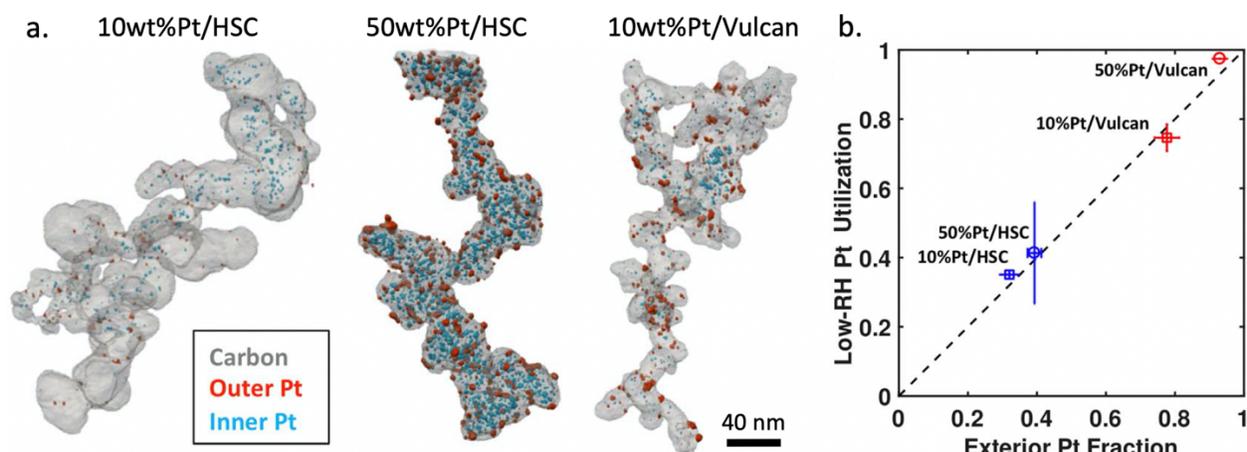

**Figure 3.** Quantitative tomographic reconstructions of fuel-cell catalyst nanostructures on different supports reveal the effects of catalyst loading (10 wt% Pt versus 50 wt% Pt) and the support structure (high-surface area carbon (HSC) versus Vulcan carbon). (a) Three-dimensional visualization showing carbon

support in gray, with Pt catalysts on the outside of the support rendered in red, and Pt catalyst particles in the interior rendered in blue. (b) At high relative humidity (RH), almost all the Pt is accessible to electrochemical utilization, but at low RH, the straight line relationship between the fraction of Pt measured on the exterior and the Pt utilization shows exterior particles dominate the catalytic activity as there is no longer a mechanism for ion transport inside the particles. Reprinted with permission from Reference 52. © 2011 American Chemical Society.

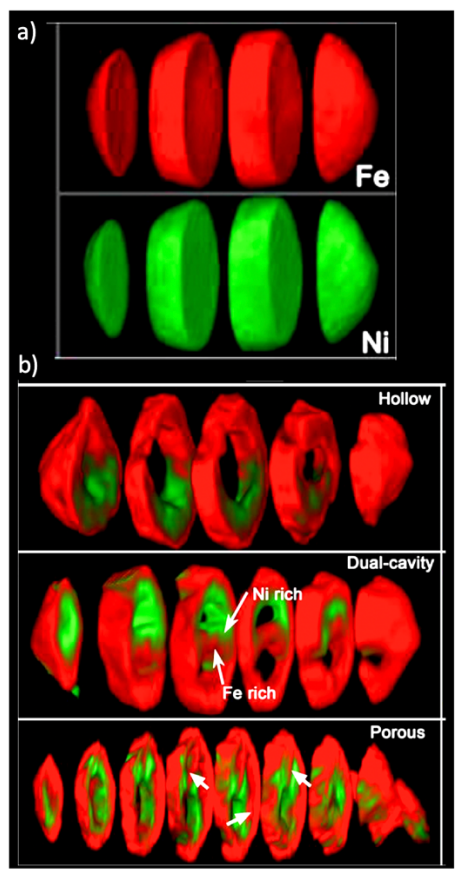

**Figure 4.** Chemical tomography of FeNi nanoparticles (a) before and (b) after full oxidation. Scale bars = 50 nm (a) and 20 nm (b). Reprinted with permission from Reference 37. © 2018 American Chemical Society.

### Three-dimensional resolution and object size

Electron tomography has a fundamental tradeoff between resolution, specimen size, and the number of projections measured, as described by Crowther.[38,39] The expression is compactly stated: $d_{3D} = \pi D/N$, where $d_{3D}$ is the resolution in three dimensions, $D$ is object size (**Figure 5**a), and $N$ is the number of projections acquired assuming equal angular spacing about a single axis of rotation. Each projection during the experiment maps to a plane of information in reciprocal space as described by the Fourier slice theorem. The missing information between each plane limits 3D resolution (Figure 5b) well before the diffraction limited resolution of the microscope $d_{2D} = 0.61\lambda/\alpha$—where $\lambda$ is the electron wavelength and $\alpha$ is the semi-convergence angle of the objective lens. For example, at 1° tilt increments, a 60-nm object has a Crowther resolution limit of 1 nm while the 2D resolution is ~1.5 Å. Thus, tomography of nanomaterials benefits little from higher resolution microscopes. In fact, higher resolution is often detrimental since the depth-of-focus diminishes rapidly and extended objects cannot be imaged entirely in focus,[40,41] the exception being atomic resolution tomography of small specimens (<10 nm)[42,43]—as discussed in a related article in this issue of *MRS Bulletin*.

Higher 3D resolution also requires substantially higher doses to achieve statistical significance in the final tomographic reconstruction. Although the dose fractionation theorem states that the total imparted dose may be divided across any number of specimen projections, it is the total dose that determines the final 3D SNR. Dose-limited 3D resolution scales inversely with the fourth root of dose ($d_{3D} \propto 1/\text{dose}^{1/4}$) and it becomes clear that the "dose required for significance is strongly resolution-dependent and 'disappointingly high' for even moderate levels of resolution," as stated by McEwen et al.[44,45] Figure 5c shows how 3D dose-limited resolution compares to the dose limits of several materials. Three-dimensional resolution for oxides is ~0.2–2 nm and ~3–10 nm for batteries based on 40–80% image contrast. In practice, image acquisition and registration require each projection to have recognizable features and a minimal dose per image. This experimentally restricts the total number of specimen tilts (e.g., 0.5–4° interval). This is especially true for chemical tomography, which has even higher dose requirements. As a result, dose limitations also impact the Crowther sampling requirements and the two become interconnected limits to 3D resolution.

The Crowther criterion ensures that all specimen features are measured and the reconstruction is free of aliasing. However, modern reconstruction algorithms attempt to estimate the missing information between specimen tilts to improve tomographic resolution. Central processing unit (CPU) and graphic processing unit (GPU) parallelization has fueled computationally demanding reconstruction algorithms that utilize iterative methods in real[46] and/or reciprocal space[47,48] and leverage prior information about the specimen.[49] A dramatic improvement to STEM tomography occurred with the introduction of compressed sensing,[50] which attempts to recover information by maximizing sparsity in a specified domain (e.g., gradient magnitude).[51]

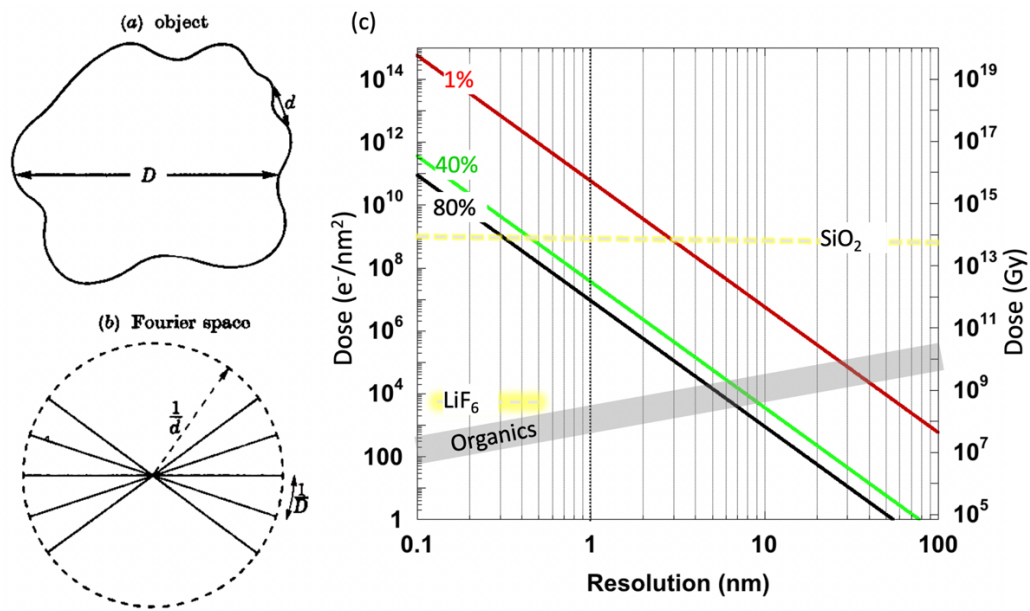

**Figure 5.** (a) The tradeoff between resolution, *d*, and object size, *D*, in electron tomography is related to the angular separation between specimen tilts. (b) This relationship is visually illustrated by the separation of missing information in Fourier space. With additional projections, larger objects and higher resolutions are achievable. Reprinted with permission from Reference 61. © 1971 The Royal Society. (c) The total dose which determines tomographic signal-to-noise also limits resolution. Three-dimensional resolution scales inversely with the fourth power of dose. Dose limited contours are plotted for 1%, 40%, and 80% image contrast.

Compressed sensing electron tomography does not measure additional specimen information, it simply provides a better estimate of the 3D object, helping to minimize artifacts from the missing information. Saghi et al. experimentally demonstrated improved reconstruction quality using compressed sensing inspired electron tomography on concave, faceted iron oxide nanoparticles (**Figure 6**).[52] Jiang et al. showed that porous materials and nanoparticles can be reconstructed reliably with fewer tilts and a limited tilt range when a sufficient amount of data is collected (e.g., 70 projections across ±70°).[53] However, compressed sensing has been demonstrated to fail when data is sufficiently undersampled,[53] or highly directional structures are present in an object.[54] With careful use, modern reconstruction algorithms can provide higher-quality 3D imaging with less information at lower doses, and research developments are continuing. However, it should be noted that data fidelity, data alignment, and sample integrity have the largest influence on reconstruction quality.

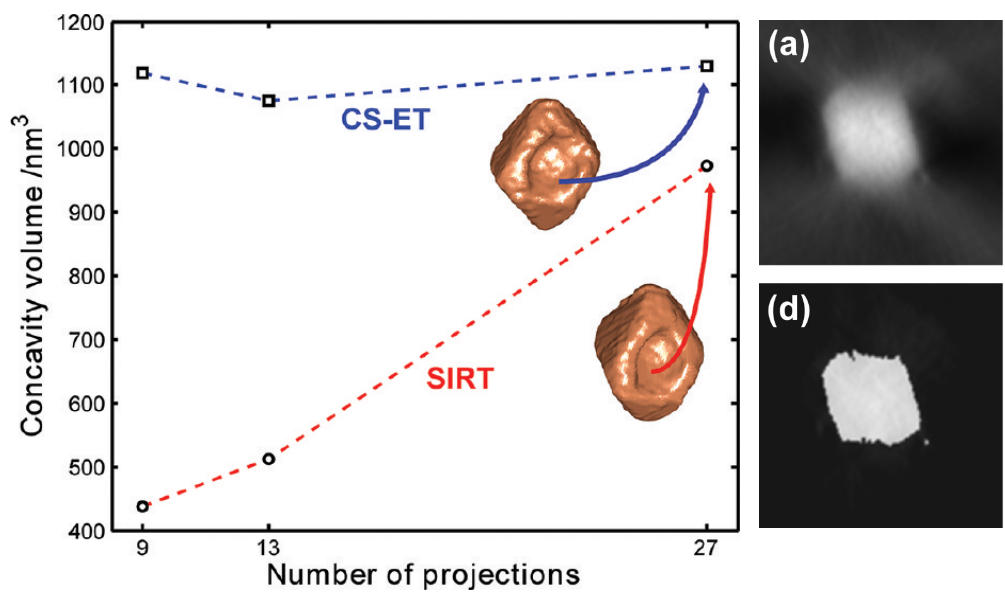

**Figure 6.** (a) Compressed sensing electron tomography (CS-ET) outperforms iterative reconstruction methods (SIRT) as illustrated from the estimation of the concavity volume of a faceted nanoparticle obtained with 9, 13, and 27 projections. The inset figures are isosurface renderings of the nanoparticle used to extract the concavity volume. (b) Cross-section from SIRT and (c) compressed sensing reconstruction. Reprinted with permission from Reference 52. © 2011 American Chemical Society.

## Going forward

Electron tomography has progressed substantially since it took hold in the 1970s. At that time, photographic plates had to be digitized and 3D visualization required physical wood blocks be cut and assembled for interpretation and analysis.[9] Today, modern instruments acquire digital data directly from multiple detectors and spectrometers with high efficiency. Software visualization tools powered by GPUs enable large volumes to be interactively rendered in three dimensions— not just as surfaces, but as transparent volumes with opacity that changes with density. While most institutions can conduct tomography on a rudimentary S/TEM, in-house expertise, appropriate software and algorithms are often unavailable. Open-source tomography tools, such as tomviz and imod,[55] as well as tutorials[56] and open data sets[1] are empowering a greater number of research groups and laboratories. As the ability to process data and visualize tomographic data increases, the use and popularity of electron tomography will continue to proliferate.

Looking forward, we can expect dose efficiency and spectroscopic tomography will improve and make new materials questions answerable, including radiation-sensitive materials such as polymers and soft-matter systems or metal–organic frameworks. Additionally, new kinds of STEM tomography will also emerge. Advanced imaging modes with arrayed pixelated detectors may enable 3D reconstruction of magnetic fields,[57–59] atomic potentials,[57] or orientational order in molecular liquid crystals and organic semiconductors.[58] The ability of pixelated detectors to record a full diffraction pattern at every probe position (leading to a five-dimensional data set) means strain fields, local crystalline order, and topological defects such as dislocations can be far more easily reconstructed in three dimensions, without requiring prior knowledge of where to sample in reciprocal space.

Reconstructing large fields of view while preserving resolution can be achieved with aberration-corrected electron tomography that combines depth sectioning.[40] Here, pixelated detectors and ptychographic algorithms could help perform the depth-sectioning without the need to record a through-focal series.[59] The industrial demand for high-throughput tomography may lead to real-time reconstruction during data acquisition. When combined with fast experimental acquisition,[60] this would allow for time-resolved electron tomography. The advancement of 3D nanoscale imaging will continue with ongoing research at the intersection of materials science, electron microscopy, and data science.

**References**
1. B.D.A. Levin, E. Padgett, C.-C. Chen, M.C. Scott, R. Xu, W. Theis, Y. Jiang, Y. Yang, C. Ophus, H. Zhang, D.-H. Ha, D. Wang, Y. Yu, H.D. Abruña, R.D. Robinson, P. Ercius, L.F. Kourkoutis, J. Miao, D.A. Muller, R. Hovden, *Sci. Data* **3**, 160041 (2016).

2. D.J. De Rosier, A. Klug, *Nature* **217** (5124), 130 (1968).

3. R.A. Crowther, L.A. Amos, J.T. Finch, D.J.D. Rosier, A. Klug, *Nature* **226** (5244), 421 (1970).

4. W. Hoppe, *Hoppe-Seyler's Z. Physiol. Chem.* **355**, 1483 (1974).

5. W. Hoppe, H.J. Schramm, M. Sturm, N. Hunsmann, J. Gaßmann, *Z. Naturforsch. A* **31** (11), 1380 (1976).

6. "Protein Data Bank in Europe," https://www.ebi.ac.uk/pdbe/emdb/statistics_emmethod.html (accessed December 2019.

7. H. Jinnai, Y. Nishikawa, R.J. Spontak, S.D. Smith, D.A. Agard, T. Hashimoto, *Phys. Rev. Lett.* **84** (3), 518 (2000).

8. A.J. Koster, U Ziese, A.J. Verkleij, A.A H Janssen, K.P. de Jong, *Three-Dimensional Transmission Electron Microscopy: A Novel Imaging and Characterization Technique with Nanometer Scale Resolution for Materials Science* **104** (40), 9368 (2000).

9. R. Henderson, P. Unwin, *Nature* **257**, 28-32 (1975).

10. T.J.A. Slater, A. Janssen, P.H.C. Camargo, M.G. Burke, N.J. Zaluzec, S.J. Haigh, *Ultramicroscopy* **162**, 61-73 (2016).

11. P.A. Midgley, M. Weyland, *Ultramicroscopy* **96** (3), 413 (2003).

12. E. Candes, J. Romberg, "Robust Signal Recovery from Incomplete Observations," *2006 IEEE Int. Conf. .Image Proc.* pp. 1281–1284 (2006).

13. L.M. Brown, P.E. Batson, N. Dellby, O.L. Krivanek, *Ultramicroscopy* **157**, 88 (2015).

14. P.A. Midgley, M. Weyland, J.M. Thomas, B.F.G. Johnson, *Chem. Commun.* (10), 907 (2001).

15. J. Radon, *Mathematische-Physische* **69**, 262 (1917).

16. H. Zhang, D.-H. Ha, R. Hovden, L.F. Kourkoutis, R.D. Robinson, *Nano Lett.* **11** (1), 188 (2010).

17. "International Technology Roadmap for Semiconductors," http://www.itrs2.net/ (2014).

18. P.D. Cherns, F. Lorut, S. Becu, C. Dupré, K. Tachi, D. Cooper, A. Chabli, T. Ernst, *AIP Conf. Proc.* **1173** (1), 290 (2009).

19. Y. Xie, W. Kim, Y. Kim, S. Kim, J. Gonsalves, M. BrightSky, C. Lam, Y. Zhu, J.J. Cha, *Adv. Mat.* **30** (9), 1705587 (2018).

20. C. Kübel, J. Kübel, S. Kujawa, J.S. Luo, H.M. Lo, J.D. Russell, *AIP Conf. Proc.* **817** (1), 223 (2006).

21. P. Ercius, M. Weyland, D.A. Muller, L.M. Gignac, *Appl. Phys. Lett.* **88** (24), 243116 (2006).

22. H.L. Xin, P. Ercius, K.J. Hughes, J.R. Engstrom, D.A. Muller, *Appl. Phys. Lett.* **96** (22), (2010).

23. G. Möbus, R.C. Doole, B.J. Inkson, *Ultramicroscopy* **96** (3), 433 (2003).

24. N.J. Zaluzec, *Micros. Today* **17** (4), 56 (2009).

25. H.S. von Harrach, P. Dona, B. Freitag, H. Soltau, A. Niculae, M. Rohde, *Microsc. Microanal.* **15** (2), 208 (2009).

26. K. Lepinay, F. Lorut, R. Pantel, T. Epicier, *Micron* **47**, 43 (2013).

27. A. Yurtsever, M. Weyland, D.A. Muller, *Appl. Phys. Lett.* **89** (15), 151920 (2006).

28. M. Weyland, P.A. Midgley, *Microsc. & Microanal.* 9 (6), 542-555 (2003).

29. M.H. Gass, K.K.K. Koziol, A.H. Windle, P.A. Midgley, *Nano Lett.* **6** (3), 376 (2006).

30. S.M. Collins, E. Ringe, M. Duchamp, Z. Saghi, R.E. Dunin-Borkowski, P.A. Midgley, *ACS Photonics* **2** (11), 1628 (2015).

31. L. Yedra, A. Eljarrat, R. Arenal, E. Pellicer, M. Cabo, A. López-Ortega, M. Estrader, J. Sort, M.D. Baró, S. Estradé, F. Peiró, *Ultramicroscopy* **122**, 12 (2012).

32. K. Jarausch, P. Thomas, D.N. Leonard, R. Twesten, C.R. Booth, *Ultramicroscopy* **109** (4), 326 (2009).

33. D. Wang, Y. Yu, H.L. Xin, R. Hovden, P. Ercius, J.A. Mundy, H. Chen, J.H. Richard, D.A. Muller, F.J. DiSalvo, H.D. Abruña, *Nano Lett.* **12** (10), 5230 (2012).

34. E. Padgett, N. Andrejevic, Z. Liu, A. Kongkanand, W. Gu, K. Moriyama, Y. Jiang, S. Kumaraguru, T.E. Moylan, R. Kukreja, D.A. Muller, *J. Electrochem. Soc.* **165** (3), F173 (2018).

35. Y. Yu, H.L. Xin, R. Hovden, D. Wang, E.D. Rus, J.A. Mundy, D.A. Muller, H.D. Abruña, *Nano Lett.* **12** (9), 4417 (2012).

36. A. Genc, L. Kovarik, M. Gu, H. Cheng, P. Plachinda, L. Pullan, B. Freitag, C. Wang, *Ultramicroscopy* **131**, 24 (2013).

37. W. Xia, Y. Yang, Q. Meng, Z. Deng, M. Gong, J. Wang, D. Wang, Y. Zhu, L. Sun, F. Xu, J. Li, H.L. Xin, *ACS Nano* **12** (8), 7866 (2018).

38. A. Klug and R.A. Crowther, *Nature* **238** (5365), 435 (1972).

39. R.N. Bracewell, A.C. Riddle, *Astrophysical Journal* **150**, 427 (1967), *Inversion of fan-beam scans in radio astronomy, The Astorphys.*

40. R. Hovden, P. Ercius, Y. Jiang, D. Wang, Y. Yu, H.D. Abruña, V. Elser, D.A. Muller, *Ultramicroscopy* **140**, 26 (2014).

41. R. Hovden, H.L. Xin, D.A. Muller, *Microsc. Microanal.* **17** (1), 75 (2010).

42. M.C. Scott, C.-C. Chen, M. Mecklenburg, C. Zhu, R. Xu, P. Ercius, U. Dahmen, B.C. Regan, J. Miao, *Nature* **483** (7390), 444 (2012).

43. Y. Yang, C.-C. Chen, M.C. Scott, C. Ophus, R. Xu, A. Pryor, L. Wu, F. Sun, W. Theis, J. Zhou, M. Eisenbach, P.R.C. Kent, R.F. Sabirianov, H. Zeng, P. Ercius, J. Miao, *Nature* **542** (7639), 75 (2017).

44. B.F. McEwen, M. Marko, C.E. Hsieh, C. Mannella, *Ultramicroscopy.*

45. M. R. Howells, T. Beetz, H. N. Chapman, C. Cui, J. M. Holton, C. J. Jacobsen, J. Kirz, E. Lima, S. Marchesini, H. Miao, D. Sayre, D. A. Shapiro, J. C. H. Spence, D. Starodub *J. Electron Spectrosc. Relat. Phenom.* **170** (1), 4 (2009).

46. P. Gilbert, *J. Theor. Biol.* **36** (1), 105 (1972).

47. Y.Z. O'Connor, J.A. Fessler, *IEEE Trans. Med. Imaging* **25** (5), 582.

48. A. Pryor, Y. Yang, A. Rana, M. Gallagher-Jones, J. Zhou, Y.H. Lo, G. Melinte, W. Chiu, J.A. Rodriguez, J. Miao, *Sci.Rep.* **7** (1), 1 (2017).

49. K.J. Batenburg, S. Bals, J. Sijbers, C. Kübel, P.A. Midgley, J.C. Hernandez, U. Kaiser, E.R. Encina, E.A. Coronado, G. Van Tendeloo, *Ultramicroscopy* **109** (6), 730 (2009).

50. E.J. Candès, T. Tao, *IEEE Transactions on Information Theory* **52** (12) 5406 (2006)

51. E.J. Candès, J. Romberg, T. Tao, *IEEE Trans. Inform. Theory* **52** (2), 489 (2006)

52. Z. Saghi, D.J. Holland, R. Leary, A. Falqui, G. Bertoni, A.J. Sederman, L.F. Gladden, P.A. Midgley, *Nano Lett.* **11** (11), 4666 (2011).

53. Y. Jiang, R. Hovden, D.A. Muller, V. Elser, *Microsc. Microanal.* **20** (6), 796 (2014).

54. J. Schwartz, Y. Jiang, Y. Wang, A. Aiello, P. Bhattacharya, H. Yuan, Z. Mi, N. Bassim, R. Hovden, *Microsc. Microanal.* **25** (3), 705 (2019).

55. J.R. Kremer, D.N. Mastronarde, J.R. McIntosh, *J. Struct. Biol.* **116** (1), 71 (1996).

56. B. Levin, Y. Jiang, E. Padgett, S. Waldon, C. Quammen, C. Harris, U. Ayachit, M. Hanwell, P. Ercius, D.A. Muller, R. Hovden, *Micros. Today* **26** (1) 12 (2018).

57. Y. Jiang, Z. Chen, Y. Han, P. Deb, H. Gao, S. Xie, P. Purohit, M.W. Tate, J. Park, S.M. Gruner, V. Elser, D.A. Muller, *Nature* **559** (7714), 343 (2018).

58. O. Panova, C. Ophus, C.J. Takacs, K.C. Bustillo, L. Balhorn, A. Salleo, N. Balsara, A.M. Minor, *Nat. Mater.* **18**, 860 (2019).

59. S. Gao, P. Wang, F. Zhang, G.T. Martinez, P.D. Nellist, X. Pan, A.I. Kirkland, *Nat. Comm.* **8** (1), 1 (2017).

60. V. Migunov, H. Ryll, X. Zhuge, M. Simson, L. Strüder, K.J. Batenburg, L. Houben, R.E. Dunin-Borkowski, *Sci. Rep.* **5** (1), 1 (2015).

61. A. Klug, *Philos. Trans. R. Soc. B Biol. Sci.* **261** (837), 173 (1971).

**Robert Hovden** is an assistant professor in the Department of Materials Science and Engineering at the University of Michigan. He received his BSc degree from the Georgia

Institute of Technology and his PhD degree in applied physics from Cornell University in 2014. His research focuses on the limits of three-dimensional imaging in electron microscopy to unveil how structure at the atomic scale and nanoscale determines macroscale material properties. Hovden can be reached by email at hovden@umich.edu.

**David Muller** is the Samuel B. Eckert Professor of Engineering in the School of Applied and Engineering Physics and co-director of the Kavli Institute for Nanoscale Science at Cornell University. He received his BSc degree from the University of Sydney, Australia, in 1991 and his PhD degree from Cornell University in 1996. He previously worked as a member of the technical staff at Bell Laboratories. His current research interests include hardware and algorithms for high-speed pixelated detectors for imaging beyond the diffraction limit, and the atomic-scale control and characterization of matter for applications in energy storage and conversion. He is a Fellow of the American Physical Society and the Microscopy Society of America, and the recipient of the Microscopy Society of America Burton Medal and Microanalysis Society Duncumb Award. Muller can be reached by email at dm24@cornell.edu.